\documentstyle{article}

\newcommand{\beq}{\begin{equation}}
\newcommand{\eeq}{\end{equation}}
\newcommand{\mn}{_{\mu\nu}}

\def\abstract#1{\vskip 7mm 
        \begin{center}{\large Abstract}\par \smallskip
                \begin{minipage}[c]{12cm}
                        \small #1
                \end{minipage}
        \end{center}
}
\def\title#1{\begin{center}{\Large\bf #1}\end{center}}
\def\author#1{\vskip 5mm \begin{center}{#1}\end{center}}
\def\address#1{\begin{center}{\it #1}\end{center}}
\makeatletter
\@ifundefined{lesssim}{}{}
\@ifundefined{gtrsim}{}{}
\def\vereq#1#2{\lower3pt\vbox{\baselineskip1.5pt \lineskip1.5pt
\ialign{$\m@th#1\hfill##\hfil$\crcr#2\crcr\sim\crcr}}}
\makeatother

\begin{document}

\title{%
  Evolution of Black Holes in Brans-Dicke Cosmology 
}
\author{%
  Nobuyuki Sakai\footnote{E-mail: sakai@ke-sci.kj.yamagata-u.ac.jp}
}
\address{%
  Department of Science, Faculty of Education, Yamagata University, 
  Yamagata 990-8560, Japan
}
\author{%
John D. Barrow\footnote{E-mail: J.D.Barrow@damtp.cam.ac.uk}
}
\address{
  DAMTP, Centre for Mathematical Science, University of Cambridge,
  Wilberforce Road, \\ Cambridge CB3 0WA, United Kingdom
}
\abstract{\
We consider a modified ``Swiss cheese'' model in the
Brans-Dicke theory, and discuss the evolution of black holes in the
expanding universe. We define the black hole radius by the Misner-Sharp mass
and find the time evolution for dust and vacuum universes. }

\section{Introduction}

The evolution of primordial black holes in scalar-tensor theories has been
studied in the literature \cite{Bar,BC,Jac,SS}. The question of what happens
to a black hole in an expanding universe in these theories was posed by
Barrow, who discussed two scenarios \cite{Bar}: (a) The effective
gravitational ``constant'' $G(t)$ at a black hole changes along with its
cosmological evolution so that the size of a black hole is approximated by $%
R=2G(t)M$. (b) $G$ remains constant at the black hole event horizon while it
evolves on larger scales; a large inhomogeneity in $G$ is therefore
generated. The case (b) was called ``gravitational memory'' because the
black hole remembers the value of $G$ at its formation time. In either case
constraints on primordial black holes would be modified \cite{BC}.

Later Jacobson claimed that there is no ``gravitational memory'' effect,
analyzing the evolution of a scalar field $\phi (t,r)$ in Schwarzschild
background \cite{Jac}. His analytic solution showed that $\phi $ at the
event horizon evolves along with its asymptotic value $\phi (t,\infty )$.
However, the solution used has a special form with very fast variation of $%
G\propto t^{-1}$ in the background universe. It was also argued that even if
the black hole mass in the Einstein frame is constant the mass in the Jordan
frame is time-dependent.

In order to investigate the time-dependence of the black hole mass, Saida
and Soda constructed a ``cell lattice'' universe in the Brans-Dicke (BD)
theory \cite{SS}. In their model the universe are tessellated by identical
polyhedrons, which are replaced by Schwarzschild-like black holes. It was
shown that the black hole mass has adiabatic time dependence, and that its
time dependence is qualitatively different according to the sign of the
curvature of the universe.

As an extension of Saida and Soda's work, we consider a ``Swiss cheese''
model \cite{DO} in the BD theory and discuss the evolution of the radius and
mass of black holes. The ordinary Swiss cheese model refers to a
cosmological model in which spherical regions in the
Friedmann-Robertson-Walker (FRW) universe are replaced by Schwarzschild
black spacetimes. Here we construct such a model in BD cosmology.

\section{Background Universe in Brans-Dicke Theory}

The BD theory is described by the action, 
\begin{equation} \label{action}
S=\int d^4 x \sqrt{-g} \left[\frac{\phi}{16\pi}{\cal R} -\frac{\omega}{
16\pi\phi}(\nabla_{\mu}\phi)^2+{\cal L}_m\right],
\end{equation}
where $\phi$ is the BD field, $\omega$ is the BD parameter, and ${\cal L}_m$
is the matter Lagrangian. The variation of Eq.(\ref{action}) with respect to 
$g_{\mu\nu}$ and $\phi$ yield the field equations: 
\begin{eqnarray}\label{EinEq}
{\cal R}_{\mu\nu}-\frac12g_{\mu\nu}{\cal R} &=&{\frac{8\pi}{\phi}}T_{\mu\nu}+
{\frac{\omega}{\phi}} \left[\nabla_{\mu}\phi\nabla_{\nu}\phi -{\frac{1}{2}}
g_{\mu\nu}(\nabla\phi)^2\right]
+\nabla_{\mu}\nabla_{\nu}\phi-g_{\mu\nu}\Box\phi,
\\\label{SFEq}
\Box\phi&=&\frac{8\pi}{2\omega+3}{\rm Tr}T.
\end{eqnarray}

As a background universe we assume the FRW spacetime:
\beq\label{frw}
ds^2=-dt^2+a(t)^2\left\{ {\frac{dr^2}{1-kr^2}}+r^2(d\theta ^2+\sin ^2\theta
d\varphi ^2)\right\},
\eeq
where $k$ is the sign of the spatial curvature. As an energy-momentum tensor
we consider a dust fluid:
\beq
T\mn=\rho u_\mu u_\nu,
\eeq
where $\rho$ and $u_\mu$ are the density and the four velocity of dust,
respectively. Then the field equations (\ref{EinEq}) and (\ref{SFEq}) reduce
the following equations for the background universe:
\beq\label{Feq1}
H^2+{\frac k{a^2}}={\frac{8\pi }{3\phi }}-H{\frac{\dot{\phi}}\phi }+{\frac
\omega 6}\Bigl({\frac{\dot{\phi}}\phi }\Bigr)^2
\eeq \beq\label{Feq2}
\ddot{\phi}+3H\dot{\phi}=-{\frac{8\pi \rho }{2\omega +3}},
\eeq
where an overdot denotes $d/dt$ and $H\equiv \dot{a}/a$ is the Hubble
parameter.

In the case of $k=0$ we know the analytic solution for the dust universe:
\beq\label{dustflat}
a(t)=a_0(t-t_{+})^{\lambda _{+}}(t-t_{-})^{\lambda_{-}},~~
\phi (t)=\phi _0(t-t_{+})^{\kappa _{+}}(t-t_{-})^{\kappa _{-}}, 
\eeq with \beq
\lambda_{\pm}={\frac{\omega +1\pm \sqrt{1+2\omega /3}}{3\omega+4}},~~
\kappa_{\pm}={\frac{1\pm 3\sqrt{1+2\omega /3}}{3\omega +4}},
\eeq
where $a_0$, $\phi _0$ and $t_{\pm}$ are arbitrary constants. If we take $
t_{+}=t_{-}$, the general solution (\ref{dustflat}) reduces to the special
power-law solution,
\beq\label{special}
a(t)=a_0(t-t_0)^{{\frac{2\omega +2}{3\omega +4}}},~~
\phi (t)=\phi _0(t-t_0)^{{\frac 2{3\omega +4}}}.
\eeq
In the limit of $t\rightarrow \infty$, the general solution (\ref{dustflat})
converges to the special power-law solution (\ref{special}). If the
present cosmic age $t$ is large enough, observations cannot constrain the
relation between $t_{+}$ and $t_{-}$. Therefore, we keep $t_{+}-t_{-}$ a
free parameter.

In vacuum case ($\rho =0$) we have the analytic solutions for all $k$. The
vacuum solution for $k=0$ is expressed as
\beq\label{vacflat}
a(t)=a_0t^{{\frac 1{3(1+\alpha )}}},~~
\phi =\phi _0t^{{\frac \alpha {1+\alpha }}}, 
\eeq with \beq
\alpha ={\frac{1\pm \sqrt{1+2\omega /3}}\omega },
\eeq
where we have omitted the arbitrary constant $t_0$ by fixing the origin of
the time coordinate $t$. Introducing the conformal time $\eta =\int {dt/a}$,
the vacuum solutions for $k=\pm 1$ are expressed as 
\begin{eqnarray}
k=+1: &a(\eta )=(\sin \eta )^{{\frac{1-\lambda }2}}(\cos \eta )^{{\frac{
1+\lambda }2}},&~~\phi (\eta )=(\tan \eta )^\lambda ,  \label{vacclose} \\
k=-1: &a(\eta )=(\sinh \eta )^{{\frac{1-\lambda }2}}(\cosh \eta )^{{\frac{
1+\lambda }2}},&~~\phi (\eta )=(\tanh \eta )^\lambda ,  \label{vacopen}
\end{eqnarray}
with \beq
\lambda =\pm {\frac 3{3+2\omega }}.
\eeq

\section{Modified Swiss-Cheese Model}

We now consider the model of a black hole embedded in the FRW universe. We
replace a sphere in the FRW spacetime with a vacuum region which contains a
black hole. Here ``vacuum'' means $T\mn=0$, and does not imply that ${\cal R}
\mn=0$ due to the existence of the BD field.

Extending Israel's junction conditions for a singular (or regular)
hypersurface \cite{Isr}, Sakai and Maeda have studied bubble dynamics in the
inflationary universe \cite{SM}. It was found that one can solve the
equations of motion for the boundary without knowing the interior metric if
the interior is vacuum, $T\mn=0$, or has only vacuum energy (a cosmological
constant), $T\mn=-\rho g\mn$. Applying this method to the present model, we
find the mass and the radius of a black hole without specifying the interior
metric, as we shall show below.

Let us consider a spherical hypersurface $\Sigma $ which divides a spacetime
into two regions, $V^{+}$ (outside) and $V^{-}$ (inside). We define a unit
space-like vector, $N_\mu ,$ which is orthogonal to $\Sigma $ and points
from $V^{-}$ to $V^{+}$. To describe the behavior of the boundary, we
introduce a Gaussian normal coordinate system, $(n,x^i)=(n,\tau ,\theta
,\varphi )$, where $\tau $ is chosen to be the proper time on the boundary.
Hereafter, we denote by $\Psi ^{\pm }$ the value of any field variable $\Psi 
$ defined on $\Sigma $ by taking limits from $V^{\pm }$.

For the matter field, we consider dust (or vacuum) for $V^{+}$ and vacuum
for $V^{-}$:
\beq\label{Tmn}
{T}{\mn}^{+}= \rho u_\mu u_\nu , ~~ {T}{\mn}^{-}= 0. 
\eeq
Although we assume a smooth boundary at which there is no surface
density, it is not obvious that this matching is possible at all times.
Therefore, we introduce the surface energy-momentum tensor on the boundary
surface,
\beq\label{Sij}
S_{ij}\equiv \lim_{\epsilon \rightarrow 0}
\int_{-\epsilon }^\epsilon dn~T_{ij}={\rm diag}(-\sigma ,~\varpi ,~\varpi),
\eeq
where $\sigma $ and $\varpi $ are the surface energy density and the surface
pressure of $\Sigma $, respectively.

Introducing the extrinsic curvature tensor of the world hypersurface $\Sigma 
$, $K_{ij}\equiv N_{i;j}$, we can write the junction conditions on $\Sigma $
as \cite{SM}
\beq\label{jc1}
[K_{ij}]^{\pm }= - {\frac{4\pi }\phi } \biggl(
S_{ij}- {\frac \omega {2\omega +3}}{\rm Tr}S\gamma_{ij}\biggr),
\eeq \beq
\label{jc2} - {\ S_i^j\ } _{|j}= [ T_i^n] ^{\pm },
\eeq \beq\label{jc3}
K_{ij}^{+} S_i^j + {\frac{2\pi }\phi }\biggl\{S_j^iS_i^j - {\frac \omega
{2\omega +3}}({\rm Tr}S)^2\biggr\}= [T_n^n]^{\pm },
\eeq
where we have defined the bracket as $[\Psi ]^{\pm }\equiv \Psi ^{+}-\Psi ^{-}$ 
and the vertical bar $|$ as the three-dimensional covariant derivative. The
junction condition for the BD field is derived from Eq.(\ref{SFEq}) as
\beq
[\phi_{,n}]^{\pm }= -{\frac{24\pi }{3+2\omega }}{\rm Tr}S, ~~
\phi ^{+}= \phi ^{-}, \label{jc4}
\eeq
which implies that $\phi $ is continuous at $\Sigma $ and inhomogeneous in $
V_{-}$.

The extrinsic curvature tensor of $\Sigma $ in the homogeneous side $V^{+}$
is expressed as \cite{SM} 
\begin{eqnarray}\label{ex1}
K_\tau ^\tau  &=&\gamma ^3{\frac{dv}{dt}}+\gamma vH,~~  \label{ex2} \\
K_\theta ^\theta  &=&{\frac{\gamma (1+vHR)}R}={\frac \epsilon R}\sqrt{
1+\left( {\frac{dR}{d\tau }}\right) ^2-R^2\left( H^2+{\frac k{a^2}}\right) },
\end{eqnarray}
where
\beq\label{defs}
R=a(t)r|_\Sigma ,~~ v\equiv a{\frac{dr}{dt}}\Big|_\Sigma , ~~
\gamma \equiv {\frac{\partial t}{\partial \tau }}\Big|_\Sigma ={
\frac 1{\sqrt{1-v^2}}},~~ {\rm and} ~~
\epsilon \equiv{\rm sign}({K_\theta^\theta })
={\rm sign}\left( {\frac{\partial R}{\partial n}}\right).
\eeq
From Eqs.(\ref{Tmn}), (\ref{Sij}), (\ref{jc2}), (\ref{jc3}), (\ref{ex1})-(
\ref{defs}), we obtain the equations of motion:
\beq\label{eom1}
{\frac{dR}{dt}} = {\frac{dr}{d\chi}}v+HR,
\eeq\beq\label{eom2}
\gamma^3{\frac{dv}{dt}}=-\gamma\biggl\{ \Bigl(1-2w\Bigr)vH-{\frac{2w}R}{\frac{dr}
{d\chi }} \biggr\}
+{\frac{2\pi \sigma }\phi } \biggl\{1+4w+{\frac{(1-2w)^2}{(2\omega +3)}}
\biggr\}-{\frac{\gamma ^2v^2\rho }\sigma },
\eeq\beq\label{eom3}
{\frac{d\sigma }{dt}}=-{\frac{2\sigma (1+w)}R}{\frac{dR}{dt}}+\gamma v\rho,
\eeq
where $w\equiv \varpi/\sigma$.

Once initial values of $R$, $v$, and $\sigma $ are given, the equations of
motion (\ref{eom1})-(\ref{eom3}) determine their evolution. As discussed in
Ref.\cite{SM}, initial values should satisfy the angular component of (\ref
{jc1}),
\beq\label{eom0}
\gamma(1+vHR)-\epsilon^{-}\sqrt{1+\left( {\frac{
dR}{d\tau }}\right) ^2-{\frac{R_{MS}}R}}=-{\frac{8\pi\sigma R}\phi}{\frac{
\omega +1+w}{2\omega +3}},
\eeq
where we choose $\epsilon ^{-}=+1$. $R_{MS}$ is defined as
\beq
R_{MS}\equiv R^{-}(1-g^{\mu \nu }R_{,\mu }^{-}R_{,\nu }^{-}),
\eeq
where $R^{-}$ is defined as $R^{-}\equiv \sqrt{g_{\theta \theta }}$ at $
\Sigma $ on the side of $V^{-}$. Because the Misner-Sharp mass is defined as 
\cite{MS}
\beq\label{MS}
M_{MS}\equiv {\frac{R^{-}}{2G}}(1-g^{\mu\nu}={R}_{,\mu}^{-}={R}_{,\nu }^{-}),
\eeq
we call $R_{MS}$ the ``Misner-Sharp radius''. Note that $R_{MS}$ is a purely
geometrical quantity and independent of theories of gravitation.

If we considered a bubble in which there is no black hole (or singularity),
we would have to solve the field equations with regularity condition at the
centre and the boundary condition (\ref{eom0}), as done in Ref.\cite{SM}.
However, because we are interested in black hole solutions, we do not have
to take such a regularity condition into account. Thus we can use Eq.(\ref
{eom0}) to determine $R_{MS}$.

At the initial time we suppose $v=0$ and $\sigma=\varpi=0$. Then $R_{MS}$
is expressed as
\beq\label{Rms}
R_{MS}={R^3}\left( H^2+{\frac k{a^2}}\right).
\eeq
Let us now discuss whether $v$ and $\sigma$ remain zero during the ensuing
evolution. Suppose $w=0$, then the only nontrivial term in Eq.(\ref{eom2})
is $\gamma ^2v^2\rho /\sigma$. If $v$ and $\sigma$ evolved from zero, 
Eq.(\ref{eom0}) shows $vH\sim\sigma /\phi$, so that $\rho v^2/\sigma\sim\rho
v/H\phi$. Therefore, Eqs.(\ref{eom2}) and (\ref{eom3}) guarantee that, if $
v=\sigma =0$ at a certain time, $v=\sigma =0$ at all time. Interestingly,
this result is true only for the dust case, $\varpi /\sigma =0$; otherwise $
(2w/R)(dr/d\chi)$ in Eq.(\ref{eom2}) would shift $v$ from zero.

In the case of Schwarzschild spacetime, the Misner-Sharp radius coincides
with the event horizon. Although it is not always true for general
spacetimes, we speculate that the Misner-Sharp radius is a well-defined
measure of the size of a black hole. In the next section, we show the
evolution of $R_{MS}$ for black holes in several background cosmological
models.

\section{Evolution of Black Holes}

The evolution of the Misner-Sharp radius for the $k=0$ dust universe is
given by Eqs.(\ref{dustflat}) and (\ref{Rms}),
\beq\label{Rdustflat}
R_{MS}=a_0^3r_0^3\left( {\frac{\lambda _{+}}{t-t_{+}}}+{\frac{\lambda _{-}}{
t-t_{-}}}\right)^2(t-t_{+})^{3\lambda _{+}}(t-t_{-})^{3\lambda _{-}}, 
\eeq
where $r_0$ is the comoving radius of the vacuum region. Equation (\ref
{Rdustflat}) shows that the black hole size decreases with time.

If we define the black hole mass as
\beq
M_{MS}\equiv {\frac{\phi R_{MS}}2},
\eeq
it coincides with the mass defined by Saida and Soda \cite{SS}. For the $k=0$
dust universe, we obtain
\beq\label{Mdustflat}
M_{MS}={\frac{a_0^3\phi _0}2}\left( {\frac{\lambda _{+}}{t-t_{+}}}+{\frac{\lambda _{-}}{t-t_{-}}}\right)
^2(t-t_{+})^{3\lambda _{+}+\kappa _{+}}(t-t_{-})^{3\lambda _{-}+\kappa_{-}}.
\eeq
It is easy to see that Eq.(\ref{Mdustflat}) reduces to $M_{MS}=$ constant,
if we choose $t_{+}=t_{-}$, which is the same result as that found by Saida
and Soda \cite{SS}. They also showed $M_{MS}$ increases for $k=+1$ and
decreases for $k=-1$, and concluded that the evolution of the mass depends
qualitatively on the sign of the curvature of the universe. We should note,
however, that their conclusion is true only for the special case $t_{+}=t_{-}
$, or equivalently, only for the asymptotic behavior of $M_{MS}$ at $
t\rightarrow\infty$.

Next, let us consider the scalar-field dominated (vacuum) universe. The
evolution of $R_{MS}$ and $M_{MS}$ in flat, open, and closed universes are
given by 
\begin{eqnarray}
k=0: &&R_{MS}={\frac{a_0^3r_0^3}{9(1+\alpha )^2}}t^{{\frac{-1-2\alpha }{
1+\alpha }}}, \\
&&M_{MS}={\frac{a_0^3r_0^3\phi _0}{18(1+\alpha )^2}}t^{-1}, \\
k=+1: &&R_{MS}={\frac{r_0^3}2}(\cos \eta )^{-{\frac{3+\lambda }2}}(\sin \eta
)^{-{\frac{1+\lambda }2}}(\cos 2\eta -\sin 2\eta -\lambda ), \\
&&M_{MS}={\frac{r_0^3}4}(\tan \eta )^\lambda (\cos \eta )^{-{\frac{3+\lambda 
}2}}(\sin \eta )^{-{\frac{1+\lambda }2}}(\cos 2\eta -\sin 2\eta -\lambda ),
\\
k=-1: &&R_{MS}={\frac{r_0^3}2}(\cosh \eta )^{-{\frac{3+\lambda }2}}(\sinh
\eta )^{-{\frac{1+\lambda }2}}(\cosh 2\eta -\sinh 2\eta -\lambda ), \\
&&M_{MS}={\frac{r_0^3}4}(\tanh \eta )^\lambda (\cosh \eta )^{-{\frac{
3+\lambda }2}}(\sinh \eta )^{-{\frac{1+\lambda }2}}(\cosh 2\eta -\sinh 2\eta
-\lambda).
\end{eqnarray}
If we take $\omega >500$, both $\alpha$ and $\lambda$ are negligible. We
see that $R_{MS}$ and $M_{MS}$ both decrease with increasing time, except in
the contracting phase of the $k=+1$ universe.

\section{Discussion}

We have constructed a modified ``Swiss cheese'' model in the Brans-Dicke
theory, and discussed the evolution of black holes for dust and vacuum
universes. We have defined the size of a black hole by the Misner-Sharp
mass, and found that it always decreases as long as the universe is in an
expanding phase.

Although we have not specified the metric around a black hole, the mass and
radius of a black hole we have obtained coincide with those of Saida and
Soda \cite{SS}, who assumed a Schwarzschild-like metric. This means that
their ansatz of the Schwarzschild-like metric does not introduce a
specialization of the problem. One fundamental problem remains. Although
the radius $R_{MS}$ represents a typical size of a black hole, the relation
between it and the event horizon has not been demonstrated explicitly and
this problem will be considered further in future work.

\vskip 5mm \noindent
{\large {\bf Acknowledgments}}\\

This work was partially supported by JSPS Programs for Research Abroad (the
1999 financial year).

\end{document}